\documentclass[natbib, smallextended]{svjour3}       
\usepackage[a4paper, vmargin=3cm, hmargin=3.1cm, marginparwidth=8cm,total={150mm,270mm}]{geometry}
\usepackage{graphicx}
\usepackage{array}
\usepackage{enumerate}
\newcolumntype{L}[1]{>{\raggedright\let\newline\\\arraybackslash\hspace{0pt}}m{#1}}
\usepackage[colorlinks,allcolors=blue]{hyperref}
\usepackage{longtable}
\usepackage{lscape} 

\usepackage{natbib}
\begin{document}

\title{Can Autism be Catered with Artificial Intelligence-Assisted Intervention Technology? A Comprehensive Survey}

\titlerunning{Can Autism be Catered with AI-Assisted Intervention Technology}        

\author{ Muhammad Shoaib Jaliawala $^{\alpha}$ $^{\zeta}$,Rizwan Ahmed Khan$^{\alpha}$ $^{\beta}$}

\authorrunning{S.Jaliawala, R.Khan} 

\institute{$^{\alpha}${Faculty of IT, Barrett Hodgson University, Karachi, Pakistan}\\ $^{\zeta}${Faculty of Engineering Sciences and Technology, Hamdard University, Karachi, Pakistan} \\
$^{\beta}${LIRIS, Universite Claude Bernard Lyon1, France.}
}

\date{Received: date / Accepted: date}

\maketitle

\begin{abstract}

This article presents an extensive literature review of technology based intervention methodologies for individuals facing Autism Spectrum Disorder (ASD). Reviewed methodologies include: contemporary Computer Aided Systems (CAS), Computer Vision Assisted Technologies (CVAT) and Virtual Reality (VR) or Artificial Intelligence (AI)-Assisted interventions. The research over the past decade has provided enough demonstrations that individuals with ASD have a strong interest in technology based interventions, which are useful in both, clinical settings as well as at home and classrooms.
 
Despite showing great promise, research in developing an advanced technology based intervention that is clinically quantitative for ASD is minimal. Moreover, the clinicians are generally not convinced about the potential of the technology based interventions due to non-empirical nature of published results. A major reason behind this lack of acceptability is that a vast majority of studies on distinct intervention methodologies do not follow any specific standard or research design. We conclude from our findings that there remains a gap between the research community of computer science, psychology and neuroscience to develop an AI assisted intervention technology for individuals suffering from ASD. Following the development of a standardized AI based intervention technology, a database needs to be developed, to devise effective AI algorithms.

\end{abstract}

\pagestyle{plain}

\section{Introduction} \label{intro}

Facial expressions play a vital role in our daily lives and our day to day social settings.  Facial expressions are an effective form of non verbal communication and provides a cue about emotional states, mindsets and intentions \citep{Kha12}. Facial expressions along with verbal cues typically let us understand what an individual really means. Human beings follow a similar pattern of development in terms of physical and mental abilities and also during the development of basic facial expressions \citep{Haa04} \citep{Fra09} \citep{Her04}. Normally developing individuals decode daily life expressions of others with whom they interact, and then behave accordingly \citep{Gro04}. Deficits in facial expressions can limit an individual's ability to network with others and impact negatively on health and quality of life\citep{Eis12} \citep{Hou88}. Deficits under certain medical conditions like ASD can also impact social skills, interactions with others as well as the perception of expressions. Similar is the case with individuals facing autism \citep{McP04}.

Neurodevelopmental disorders such as ASD are linked with abridged ability to produce \citep{McI06} and perceive \citep{Ado01} facial expressions \citep{Rum09}. Defining criteria for autistic disorder, as set by the diagnostic handbooks, and accepted worldwide such as ICD-10 WHO (World Health Organization 1992) \citep{Org93} \& DSM-IV APA (American Psychiatric Association) \citep{Ass00} are: abnormalities in social interaction, verbal and non-verbal communication impairments and a limited range of interests and activities. Verbal social impairments include deficit in social interactions such as dealing with everyday conflicting scenarios \citep{Ber01}, communicating with people \citep{Ber01}, dealing with people in the society, language and reading skills, spontaneous greetings to peers. While non-verbal communication impairments include deficit in facial expressions, physiological changes \citep{Ros12}, maintaining eye contact and decision making. To cater these deficits and disruptions of recognition of facial expressions, computer scientists are trying to enhance the skills and perception of these individuals facing ASD, by technology, such as Computer Aided Systems (CAS) and Computer Vision Assisted Technology (CVAT) utilized in making serious games \citep{Tan10} \citep{Tsa11} \citep{Jai12} \citep{Coc08} and making different skill set batteries \citep{Tan10} \citep{Bar04} \citep{Sil01} \citep{Wha10} \citep{Str13}. For discussion on different pervasive developmental disorders (PDD) including autism, refer Section \ref{BDASD}.

According to the British Machine Vision Association (BMVA), computer vision is the science that aims to give machines their eyes and brains to see and visualize the world. It is concerned with the automatic analysis, extraction and understanding of information from the data in the form of biometrics\citep{Lim01}, character recognition \citep{Gov90}, forensics \citep{Joh05}, image restoration \citep{Kha13}, medical image analysis \citep{McI96} as well as facial expression recognition \citep{Kha12a}. It involves the development of a theoretical and algorithmic basis to get automatic visual understanding.


Researchers studied and utilized automatic expression recognition algorithms  to aid individuals having ASD by intervening in social skills, expression recognition and production of expressions e.g by serious games  \citep{Jai12} \citep{Coc08} \citep{Sil01} \citep{Bea08}. It has been theorized that technology is independent from inclination towards someone depicting maladaptive behaviors and is just, whoever is using it. While humans (therapists) tend to reinforce maladaptive behaviors unintentionally \citep{Pli85} causing the results to be not concrete or not exactly true to make conclusions. Maladaptive behaviors are behaviors that are commonly and most frequently used to reduce an individual's anxiety, but the results are dysfunctional and non-productive. For instance, avoiding some scenarios because you have unrealistic fears may for the time being reduce your anxiety, but it is non-productive in alleviating the actual problem in the long term. 

Up until now there is no universally accepted treatment, intervention or cure for ASD \citep{Rog98} \citep{Coh06}. But there have been numerous cases reported where intensive behavioral and educational intervention programs significantly improve outcomes for individuals in the long run \citep{Rog98} \citep{Coh06}. To the best of our knowledge, no empirical generalized results have been concretely reported yet, although there are studies which demonstrate positive and convincing outcomes in post intervention results \citep{Gry14}. There is an urgent need for intensive treatment methods for individuals facing autism. But appropriate interventions, resources and therapeutic treatments are difficult to access or are just too expensive \citep{Tar02}. For this reason, CAS are being utilized to create automated and easily accessible systems to cater the needs of the individuals having ASD.

Some previously done research surveys on ASD \citep{aresti2014technologies}, \citep{Ram11} give evidence that CAS treatments are more robust on those individuals that do not face severe mental abnormalities, like HFA (High Functioning Autism) and AS (Asperger's Syndrome). Refer to Table \ref{table:ASDtable1} to read about HFA, AS and other related developmental disorder sub-categories.

In this survey, we present a consolidated literature related to the assistance of individuals with ASD having impairments in facial expressions (perception, production, enhancement and encouragement). Besides discussing the work, we also discuss CAS / CVAT developed to enhance the capabilities of individuals having ASD. The term ASD (in this survey) is used for all the individuals on the autistic spectrum. Most of the currently available literature and research findings are based on individuals with HFA or AS, except few on  LFA (Low Functioning Autism) \citep{Hil03}. Also, the term \textit{Emotions} is used to refer to only the facial expressions produced e.g.( happiness, sadness , surprise, etc) in response of some emotional state.


The rest  of  the  article  is  organized  as  follows: details related to autism are defined in Section \ref{BDASD}. Literature related to the distinctive CAS / CVAT / Virtual Reality (VR) based interventions is presented in Section \ref{FEBI}. Section \ref{Sum} discusses a summary of the literature in tabular form followed by the conclusion and future directions.

\section{Brief Description of Persuasive Developmental Disorders (PDD)} \label{BDASD}


\begin{table*}[!htbp]
\centering
\begin{tabular}{||p{3cm}||p{3.2cm}||p{6.5cm}||}

 \hline
& & \\		&  & Qualitative impairment in social interaction, communication (delay or lack of development of spoken language) and restricted, repetitive and stereotyped patterns of behavior, interests, and activities categorized by substantial difficulties in social interaction, non-verbal communication skills  \\ \cline{3-3} & & \\ & Autistic Disorder &  High Functioning Autism (HFA) is a term applied to people with autistic disorder who are deemed to be cognitively ``higher functioning'' (with an IQ of 70 or greater) than other people with autism. Individuals with HFA may exhibit deficits in areas of communication, emotion recognition and expression, and social interaction.  
\\ \cline{3-3} 
& & \\ && Low Functioning Autism (LFA) refers to autistic people with cognitive impairments. Symptoms may include impaired social communications or interactions, bizarre behavior, and lack of social or emotional reciprocity. Sleep problems, aggressiveness, and self-injurious behavior are also possible frequent occurrences\\ \cline{3-3}\cline{2-2}
												
												 & &\\ & &\\											
 
& Asperger's Syndrome (AS)  
& Qualitative impairment in social interaction, restricted repetitive and stereotyped patterns of behavior, interests, and activities; no clinically significant general delay in language or cognitive development. 
Generally have higher IQ levels but lack in facial actions and social communication. 
  \\ & &\\  \cline{3-3} \cline{2-2}
 & &\\
 \bfseries Pervasive Developmental Disorders  (PDD)  & Rett's Disorder &	Development of multiple specific deficits following a period of normal functioning after birth. Deceleration of head growth, loss of previously acquired purposeful hand skills, loss of social engagement early in the course of life.
\\
	& 
	&Appearance of poorly coordinated gait or trunk movements. Severely impaired expressive and receptive language development with severe psycho-motor retardation.
 \\ \cline{3-3} \cline{2-2}
	 & & \\  & Childhood Disintegrative Disorder & Marked regression in multiple areas of functioning following a period of at least two years of apparently normal development. Expressive or receptive language; social skills or adaptive behavior; bowel or bladder control; or play or motor skills  \\ \cline{3-3} \cline{2-2}
	& Pervasive Developmental Disorder Not Otherwise Specified (PDD-NOS) &Severe and pervasive impairment in the development of reciprocal social interaction or verbal and non-verbal communication skills and stereotyped behaviors, interests, and activities.The criteria for autistic disorder are not met because of late age onset; atypical and/or sub-threshold symptomotology are present.

 \\ \cline{2-2} \hline

\end{tabular}
\caption{Sub-categories of Pervasive Developmental Disorders (PDD). This table is inspired by literature presented in the DSM-IV and studies reported by Autism Society of America (ASA).} \label{table:ASDtable1}
\end{table*}


Autism is defined using behavioral criteria, since so far no genetic markers are known. The characteristics vary considerably in severity as well as in combination, within and across individuals, as well as with time \citep{Ass00}. Apart from AS or HFA; the subgroups of autism are the terminologies used to describe the higher functioning end of the autism spectrum. AS correspondingly with HFA is a newer term as it was recognized later on, in early 1994 \citep{Ass00} \citep{Ozo00}. Currently, all the terms have been merged under a single umbrella called Autism Spectrum Disorder (ASD). Since we have classified and surveyed all the research since 1973 all the terms are mentioned categorically according to the DSM-IV criterion and the Autism Society of America. For a clearer understanding of different sub-categories of developmental disorders refer to Table \ref{table:ASDtable1}.

Due to above mentioned associations and the extension in spectrum, the criteria has led to a dramatic increase in the diagnosis of ASD. Thus, autism is not anymore a rare disorder \citep{Hil03}. AS includes individuals who have fluent language and better academic skills along with obsessions and narrow interests, though they have limited facial expressions and damaged recognition as well as interpretation of emotional perceptions \citep{Fit04}. Children having ASD face problems associated with communication and usually misinterpret by depending on literal, rather than the contextual meaning of words \citep{Gry08}. Such children also face other problems such as tantrums and self-injurious associations \citep{Ber01} \citep{Plo10}. All these criteria has been agreed upon around the world by researchers \& clinical practitioners.

Population statistics depict that autism is a rapidly emerging developmental disability in the United States, with an estimated annual expense for diagnosis, and treatment of 90 billion USD (Autism Society of America) which rose up to 236-262 billion US dollars in 2014 \citep{Bue14}. Statistics also reveal that about 1\% of the world's total population has Autism Spectrum Disorder, as reported by the Center for Disease Control and Prevention (CDC) \citep{Ric09}. The total population ratio of Male:Female is 3:1 \citep{Cha01} \citep{Bai00}. This proves that autism majorly and mostly occurs in males. It has also been stated that mental abnormality, which means IQ less than 70 is strongly associated with autism and is in between 25\% to 40\% cases of ASD \citep{Cha01} \citep{Bai00}. Comparably, Aspergers Syndrome(AS) is estimated to affect at an even higher Male:Female ratio, ranging from  4:1 to 10:1 \citep{Cha01} \citep{Bai00}.

The use of CAS / CVAT is being utilized by researchers and is comparatively an inexpensive alternative for people and therapists.
 The utilization of computer science in collaboration with psychology in the treatment and studying of autism for research purposes was recognized early on \citep{Col73} \citep{Mat89}. However, only since the last decade researchers have been investigating the application of CAS mapped with Autism\citep{Wer01} more rigorously.

\section{Distinct CAS / CVAT Based Interventions}\label{FEBI}

A question from the researchers that is being handled quite critically, is whether CAS actually is more effective than traditional teaching methodologies or not. Researchers have reported positive outcomes  but only in terms of statistical inferences \citep{Gry14}. Some notable studies have specifically addressed the subject of the efficacy of CAS over traditional approaches \citep{Osp08} with individuals facing ASD \citep{Wha10}\citep{Pen10}.

Regardless of whether CAS is better than traditional approaches in terms of efficiency or not, CAS can be given preference over traditional methods just because it could be more easier with CAS to devise treatments with unlimited repeats, greater precision and lesser variability thus,  ensuring higher treatment fidelity. This would also make it possible to reach-out remote areas at a larger scale by saving cost, due to automation. Along with reduced requirement for highly qualified and trained, expensive, service providing professionals, this also paves way for larger rate of diffusion of treatment, training, and education. We can also discern that CAS compared to human interventions is more effective because the  helper or the attendant in human instructed sessions can induce maladaptive behaviors unintentionally while giving individual more attention.

In the facial expression based interventions we have sub divided the categories into interventions that target 

\begin{itemize}
\item Non-Verbal Communication Skills (refer to Section \ref{NVCS}) i.e.  facial expressions, emotion recognition, affect recognition, maintaining eye contact, decision making, etc.

\item Verbal Social Skills (refer to Section \ref{VSS}) i.e. social interaction such as dealing with everyday conflicting scenarios, communicating with people, dealing with people in the society, language and reading skills, spontaneous greetings to peers, etc.

\item Virtual Reality / Augmented Reality (refer to Section \ref{VR}) i.e. technology inculcating both verbal social skills \& non-verbal communication skills in an Augmented Reality based CAS / CVAT intervention.
\end{itemize}

\subsection{Non-Verbal Skills} \label{NVCS}
Communication with people is much more than what a person has to say or the messages given by physiological changes and facial expressions. This includes implicit messages whether intentionally or unintentionally which are expressed through non-verbal cues.
 Non-verbal communication includes facial expressions, the tone and pitch of the voice, gestures, body language (kinesics) and distance between the communicators (proxemics) but the main focus lies with the expressions of the face.

Plienis et al. \citep{Pli85} reported seventeen individuals (aged 4-14 years with mean age 8 years 2 months) out of which 6 individuals had ASD. Individuals were involved in both human and computer instructed sessions in an alternating treatment design which consisted of judgmental tasks. The displaying video was faced by a two key plexiglass response panel that seperated the screen down the vertical midline.The apparatus was situated in a medium sized sound diminished room illuminated by florescent bulbs (2x 15W). Plienis et al. \citep{Pli85} theorized that CAS sessions are equally efficacious to human instructed methods which was contrary to the previous studies \citep{Richmond1983} \citep{Russo1978} but according to the claims of the authors the reason for this efficacy was high quality of the speech synthesizer, developed / programmed to produce verbal instructions for the individuals. 

Chen et al. \citep{Che93} compared computer and human-instructed tasks with four individuals age ranging between  4-7 years with ASD, to understand independently selected tasks. They developed an enthusiasm scale to assess whether CAS was more amusing than traditional human instructed mechanism. Moreover assessing whether CAS resulted in a better task performance comparatively. Like Plienis et al. \citep{Pli85}, Chen et al. \citep{Che93} theorized enhanced motivation in individuals using CAS.

Kodak et al. \citep{Kod11} performed experiment with only one female individual (age 7 years) which showed positive results in showcasing the supremacy of CAS over human instructions. There were several limitations in the experiment performed by Kodak et al. \citep{Kod11}, for example inclusion of a therapist to move the CAS to the next trial. Secondly, the CAS was not designed to assess generalization of the targets in a natural environment. 

Whalen et al. \citep{Wha06} introduced \textit{``TeachTown''} a self-paced lessons and reward game, which incorporated principles of Applied Behavior Analysis (ABA) for correct responses. The lessons focused on domains like cognitive skills, social \& life skills as well as receptive language skills. Whalen et al. \citep{Wha10} studied individuals with ASD specifically to measure the efficacy of TeachTown: Basics- an educational CAS that included computer lessons and day to day environmental activities that involve principles of ABA. The individuals were trained in a distinct trial format with a reinforcement on correct responses in an (timed) animated game followed by verbal praise and graphics, alongside human instructed approach. Individuals received daily sessions on the CAS for twenty minutes approximately and a same physical activity session as part of the TeachTown CAS protocol. Total 47 childhood individuals with ASD aged 3-6 years were involved. The study reported improvements in receptive language and social skills after 8 weeks of training with ``TeachTown''-CAS. Several limitations were reported as some instructors did not follow the proper curriculum, data was not collected from a certain period of time as it was not auto logged, and very minimal time was spent on the CAS.

Researchers in the early millennium \citep{Sil01} worked on improvement and comprehension of emotions in individuals with ASD and AS with the \textit{``Emotion Trainer''} CAS. A single group comprising of 11 individuals (children) received CAS training while the other group (control group) attended as a non-CAS (i.e. did not got the intervention) comprising of 11 individuals (children). The CAS can be easily understood by compartmentalizing it into 5 sections: the first section is to identify emotions (e.g. sad, happy, angry) from a facial expression; described and displayed in Howlin et al. \citep{How99}. The second section showed cartoon characters with a label describing a scenario. The scenario would prompt an emotional expression e.g (a rabbit's image, labeled ``Mary's pet rabbit died'') and prompted whether the scenario would make Mary happy, sad or afraid. The third section displayed an image of what a person desires and what expressions would there be if they don't get what they desire or what if they receive something else apart from what they desire e.g. ``Josh wants a pizza, but gets a sandwich.'' Now whether that made Josh happy or sad. The fourth section is similar to the second section but addressed mental states rather than the physical events. Section 5 displayed an incident and asked whether it was liked or disliked. Throughout the sections, accurate responses were awarded by compliments "well done" displayed on the screen. \citep{Sil01} reported significant statistical improvement in section 2 and section 4 of the CAS. Though no significant improvements for facial expressions were reported. 

B{\"o}lte et al. \citep{Boe02} designed and evaluated a CAS named FEFA to train basic facial expression recognition skills to individuals with ASD. Study utilized an experiment of 10 (male) individuals with HFA/AS, aged 16-40 years. The participants were randomly divided into 2 groups, N = 5 to a group that would take the CAS intervention. While the other control group, N=5 were assessed without intervention. 1000 photographs of properly distinguishable facial expressions of females and males were collected from people of different cultures. These included the pictures of facial affect from \citep{Ekm72}. The images were cropped down just to show the eyes rather than the complete face for recognition of expression for a particular modality. For rating the photographs, cross-cultural concept of universal basic emotions \citep{Ekm71} was chosen as reference. Two tests were designed, first in which a complete face with an expression representation with options of what is the person feeling at the moment. While the other showed a cropped image of the eyes with the same options of how was the person feeling. For correct answer, a feedback was achieved with smiles next to the correct answer. If a wrong answer was given, a feedback sign would be displayed. If the individual selected the wrong answer, the right answer will be displayed along with the explanation. The comics used were from \textit{``Teaching children with autism how to mindread''} \citep{How99}. However, the engagement of CAS with individuals having ASD may need to be taught specifically \citep{Kag11}. \citep{Boe02} believe that the training module have not significantly contributed towards the intended behavior modification in everyday life as expected by the authors. 

B{\"o}lte et al. \citep{Boe06} also reported that the effect of CAS is to instill facial affect recognition while observing neurological structures by functional Magnetic Resonance Imaging (fMRI) of individuals with ASD. 10 participants were included in the study out of which 5 were control group and the other 5 were in the intervention group (mean age 29.4 years). The fMRI scans were made before and after the experimental group had gone through the CAS intervention phase. FEFA CAS \citep{Boe02} was used to train and assess  the recognition of facial expressions (basic emotions).  No significant activation changes in the fusiform gyrus (part of the temporal lobe and occipital lobe - brain) was observed via fMRI. Though trained participants showed behavioral improvements while considerable improvement in the basic detection of emotional skills was also reported. As limited individuals were present for the study so no generalization could be made and this was reported as a limitation as well. Moreover there was no empirical data to show the clinical significance and generalization of these effects.


Moore et al. \citep{Moo05} developed a CAS as visual representations of facial expressions with an ``avatar''. The CAS allowed representation of facial expressions and animated sequences conveying emotional scenarios. The avatar displayed four exclusive expressions: happy, angry, sad and fear. The study was conducted in three different stages. In the first stage, the individuals were asked to recognize expression of the avatar, by choosing the corresponding reply-representation. In stage two, the individuals predicted the expression in different scenarios.  While in stage three, individuals were shown an avatar in a scenario expressing an emotional expression and were asked to choose which event(s) caused the expression. Analysis of the data was done by relatively comparing the total observed responses to the total responses expected if they were selected by chance. The author reported that 90\% of the individuals who participated, were able to recognize, predict and interpret the avatar's emotional expressions. Moreover, they concluded that this CAS methodology allowed the children to effectively communicate with others.

Golan et al. \citep{Gol06} did two experiments in which individuals were trained with the \textit{``Cambridge MindReading''} CAS to recognize expressions in other human faces. The core idea of the CAS, was facial expression recognition and audio expression recognition after showing or playing a facial expression video of 3-5 seconds or auditory clip of an emotional intonation in words. After watching the facial expression clip or the audio, the participants would be given 4 adjectives and asked to ``choose the word that best describes how the person is feeling''. To ensure that the concepts were selected from the adult emotional range, they were selected from the higher levels of the taxonomy; six from level four (concepts which were understood easily by a typical 15-16 years of age group), thirteen from level five (concepts which were understood easily by a typical 17-18 years of age group) and one from level six (concepts which were understood easily by  less than 75\% of typical 17-18 years of age group). The model also comprised of 412 unique concepts, which  were clustered into 24 mutually exclusive clusters such as angry group, happy, sorry group, etc. While using these taxonomies along with visuals and auditory enhancements, Baron et al. created a database of actors of both sexes of different age and ethnicities. The ``MindReading'' CAS addresses to assess the wide range of emotional expressions and to thoroughly measure both facial and auditory modalities. A more detailed information can be obtained from \citep{Bar04}. In Golan et al. \citep{Gol06} experiment, a total of 65 individuals participated. Out of 65 individuals / participants, 41 participants (mean age of 30.5 years) were facing HFA / AS \& having a verbal IQ \& performance IQ of 108.3 and 112 respectively. 24 experiment participants (mean age of 25.3 years) were typically developing individuals having verbal IQ 115.8 and performance IQ of 112.5. In the first experiment HFA / AS individuals were randomly divided into two groups. 19 individuals along with the 24 typically developing individuals went through the CAS intervention at home for 10-15 weeks. While the other 22 individuals were assessed without any intervention i.e. a control group. Experiment 2 repeated the same structure / procedure but with added weekly support of a tutor along with the CAS intervention. The comparison between intervention and non-intervention of ``MindReading'' CAS was conducted along with the comparison of AS / HFA individuals across 3 generalization levels. \textit{close generalization, feature-based distant generalization and holistic distant generalization} \citep{Gol06}. Participants showed no significant improvements in the scores of the CAS at Time1 (T1) i.e. pre CAS intervention, apart from the typically developing individuals that performed better than the AS / HFA individuals at all recognition tasks across the \textit{Holm's sequential rejective Bonferroni procedure} \citep{Hol79} \citep{Zha97}.
 
 
Lacava et al. \citep{Lac07} also used \textit{``MindReading''} CAS to teach eight individuals (children aged between 8-11 years) with Aspergers Syndrome(AS). Out of eight children, six were males and two were females. Lacava et al. reported significant improvements in children's ability to identify basic to complex emotions. Another positive aspect reported was, the individuals participating in the experiment found the CAS more enjoyable and entertaining rather than the traditional and conventional methods. Individuals showed improvement in operating the CAS. In an another experiment, Lacava et al. \citep{Lac10} used ``MindReading'' CAS to teach four individuals with HFA to identify the basic to complex emotions resulting in positive outcomes i.e. improvements in the recognition but no concrete conclusions were made. The constraint of \citep{Lac07}\citep{Lac10} was the limited and small sample set / size of individuals taking part in experiment thus generalizations on the results cannot be made.

Faja et al. \citep{Faj07} found that CAS based face training can affect processing of faces. Faja et al. observed that 7-10 hours of CAS based facial training results in sensitivity to holistic face processing. In Faja's experiment five young adult males with HFA-ASD having mean age of 19.0 years and a full scale IQ of 99.0; were given training to identify faces according to sex, chronological age and identity. After 3 weeks of training, post-intervention results revealed that the trained group demonstrated more sensitivity to configurable information i.e. distance in between the eyes relatively to the untrained lot. This depicted that identity recognition skills can be improved through practice in recognition of the face. However, this approach had several limitations like small sample size, diverse age range and inclusion of only individuals with HFA. We conclude that generalizations cannot be made for treatment of face processing from \citep{Faj07} as sufficient information is not available.

Beaumont \& Sofronoff \citep{Bea08} compared a treatment group and a control group measured on facial expressions, social competence and understanding of emotional management strategies.  Beaumont et al. developed a CAS called the \textit{``junior detective training''} which comprised of three levels. First level taught participants to decipher facial expressions along with body postures, as well as prosody of speech of a CAS managed human like character. The second level worked with individuals to decode the emotions of cartoonic characters in variety of scenarios using non-verbal cues. The third level provided the individuals to use those learned social skills to be utilized in a Virtual Reality (VR) environment e.g. dealing with a bully. Post treatment results depicted statistical improvements comparatively from the control group. Two statistical measuring reports were concluded: parent's report and an instructor's report. The reports were questionable because it could not be assessed whether they proved clinical perspective.

Tanaka et al. \citep{Tan10} assessed facial expression processing with a curriculum-based, experimental measure called the \textit{``Let's Face It!''} (LFI) CAS \citep{Wol08}. The proposed measure was composed of seven sub-tests (game modes) that were based on the perception of facial identity over a wide range of face processing tasks defined by \citep{Wol08}. The seven modes were: parts matching to complete faces; distinguish between changes in face dimensions; recalling / memorizing faces; face matching with masked features; face matching of same individual with different expressions; distinguish between changes in dimensions of a house and recalling / memorizing cars. A total of 117 individuals / participants facing ASD, from which 79 children, adolescents and young adults were included in the study, and were randomly divided into two groups, active treatment and a wait-list group. The active treatment group consisted of 42 participants (34 males and 8 females) were analyzed out of 65 with a mean age of 10.5 and a mean full scale IQ of 93.6 while the wait-list group had 52 participants out of which 37 (28 males and 9 females) were analyzed  with a mean age of 11.4 years and a mean full scale IQ of 95.9 . The most significant sub-test reported was to be parts / Whole identity i.e. parts matching to complete faces based upon the ``Let’s Face It!'' CAS \citep{Wol08}. Participants outcomes for treatment differed significantly only on the test for matching parts to whole faces. For further study, review \citep{Wol08}.

Cockburn et al. \citep{Coc08} developed \textit{``SmileMaze''}, an expression production recognition CVAT to enhance skills  of children dynamically, and made engaging by a game format. The Computer Expression Recognition Toolbox (CERT) \citep{Bar08} is the backbone framework used for the ``SmileMaze''. It is similiar to a Pacman game \citep{Khe13} (look wise) while the rules are different from a Pacman game. The user or player is a pacman (sprite) which is navigated by the keyboard navigation keys while other, expression imposed sprites, that act as barriers in the pathways of the maze are removed by mimicking the expression imposed on the barrier sprites (i.e. happy / smiling). No interaction, study or testing whatsoever of the system with individuals of ASD was reported. Cockburn et al. also compared their CVAT with \textit{``lets face it!''} (LFI) CAS \citep{Wol08}. But no meaningful comparison was done as the LFI only provides recognition and not the production of facial expressions.

Golan et al. \citep{Gol09} developed an animated series named \textit{``The Transporters''} to enhance the facial expressions as well as emotional vocabulary. The Transporters is based on eight characters who are vehicles that move by a rule based approach. While on the vehicles were real human faces of actors showing different facial expressions which were super imposed. These vehicles were then utilized in a thematic short storyline scenario e.g ``Charlie is feeling happy with a vehicle having a face of a human showing happiness, after that showing other vehicles having faces, to select who else is feeling happy ?''. Golan et al. based their study onto the Empathizing-Systemizing (E-S) theory \citep{Bar09}.

Hopkins et al. \citep{Hop11} tested their intervention named \textit{``FaceSay''} CAS. ``FaceSay'' contains three distinct modalities with avatars designed to teach social skills to individuals having ASD. The avatars were animated photos of real persons while the motto of the games was to promote eye gaze, facial expressions recognition and face recognition. The first modality was to attend to eye gaze. The avatar was surrounded by different objects. The player was asked to touch the object where the avatar is gazing towards. Correct responses display appreciation. While the second modality was designed to teach face recognition. The player was asked to select the proper fit for that distortion present on the avatar's face. While the third modality was to teach to perceive the eye movements. The study was carried out with a group of HFA and LFA. 49 Individuals (24 HFA \& 25 LFA) having IQ of 91.9 and 55.1 respectively were involved in ``faceSay'' CAS intervention that targeted eye gaze and facial expressions and emotions. Individuals with HFA demonstrated successful emotional recognition and social interaction. The LFA group didn't show promising or significant improvements holistically.

Tsai et al. \citep{Tsa11} used CVAT named \textit{``FaceFlower''} to aid ASD individuals to improve non-verbal (facial expressions) by using a flower to grow by the user's (individuals with ASD) facial expressions. The motive was to aid individuals to pronounce their facial expressions and practice their expressions and facial muscles to a particular expression in order to see the plant grow.  The CVAT was controlled through the facial expressions of the individual. The facial expression detection tracking and feature extraction was done using \textsl{eMotion} software that classifies the expressions based on facial emotions defined by FACS \citep{Ekm78}. eMotion is a tool for creating interactive motions of objects for visual performances. However there wasn't any experiment done on any sample size through which results could be concluded.

The Facial Action Coding System (FACS) \citep{Ekm78} is a detailed structurally  based arrangement for measuring all visually apparent facial movements. FACS defines all visually unique facial activity based on 44 unique Action Units (AUs) along with multiple classifications of head and eye placements and movements. FACS procedures also allow for the coding of the intensities of each facial action on a scale of 1 to 5 point intensity. For a detailed list of all AUs and procedures review \citep{Ekm97}. 

Faja et al. \citep{Faj12} studied the expertise in effect of training faces in adults facing ASD and those showed initial impairments in face recognition. The authors reported the results of study in which the individuals who participated were randomly grouped into two. First group was to take computer training involving faces and recognizing the attributes about that face such as gender. The other group was shown houses and asked to describe the shape of the house. The authors evaluated that participants in face recognition and participants in house recognition, both showed improvements on measures of memorization of the faces and houses.

Jain et al. \citep{Jai12} proposed a CVAT for ASD which incorporates domains of computer-vision and computer graphics. The CVAT tracks facial features and utilizes those features in the recognition of facial expressions of individuals. The system was built to recognize six basic / universal expressions \citep{Ekm79} i.e. \textit{anger, disgust, fear, joy, sadness and surprise}. These tracked features were also used to animate an avatar which eventually mimics individual's expressions. This was accomplished by locating and tracking facial features, approach detailed in \citep{Jai11} .



\subsection{Verbal Social Skills} \label{VSS}

Verbal social skills are defined as there is a high cohesion of verbal intonation and non verbal facial expressions while communicating. For instance, when a person is fearful, his voice trembles and the fear can be seen by their facial expressions as well as their voice. Though the CAS that was designed specifically for audio or voice training or teaching communication are not part of this review. But the facial expression CAS training or analysis that has voice training and also caters to the verbal social skills are included and discussed.

Social skills are the skills we utilize regularly to interact and speak with others. Social skills are essential in empowering a person to have and keep up smooth interactions. A considerable lot of these abilities are critical in making and maintaining relations and fellowships. Social interactions don't always run smoothly but an individual's needs to execute appropriate strategies, such as settling clashes when problems in interaction come up. It is likewise imperative for people to have ``empathy'' (i.e. having the capacity to place yourself into another person's shoes and perceive their sentiments) as it enables them to react in an understanding and minding approach to how others are feeling.

Simpson et al. \citep{Sim04} designed a CAS to instill social skills to fourth grade-schoolers aged 5-6 years. All individuals had deficiency of social skills and were facing mild to severe, language and speech impairment. \citep{Sim04} designed a CAS consisting of video-clips of typically developing individuals depicting some instances of target behavior like sharing, following instructions of a teacher and social greetings. The authors also assessed the grade-schoolers on the contrary in vivo generalization sessions. In these sessions, the individuals typically interacted with control developing children. The post intervention gave improvements in the targeted social skills though specific component analysis was not possible as they employed a multi-probe design across all the behaviors. Thus, authors cannot assess which particular aspect of treatment was responsible for the improvements.

Whalen et al. \citep{Wha06} introduced \textit{``TeachTown''} a self-paced lessons and reward game, which incorporates principles of applied behavior analysis for correct responses. The lessons focused on domains like cognitive skills, social \& life skills as well as receptive language skills. Nikopoulos et al.\citep{Nik07} used video modeling in two experiments as CAS involving a couple of baseline designs. Four individuals with Autism were taught social sequences incorporating social initiation reciprocal play, imitative responses, and object engagement. Videotapes were shown to individuals with respect to the social skills acting as a marker for each target behavior. Every individual watched all the video-clips and were free to go in a place where individuals were confronted by a group of people for social interactions. The authors reported positive results evidencing video modeling to be fruitful and successfully helped individuals with ASD.

Lacava et al.\citep{Lac07} used \textit{``mindreading''} CAS to teach eight individuals (children) with AS having age 8-11 years (6 males and 2 females) to identify the basic to complex emotions. The authors concluded significant improvements in verbal social skills and non-verbal social skills. Another beneficial effect that was reported was, the individuals participating found the CAS or computer intervention more fun and entertaining rather than the traditional and conventional ones and showed improved skills in operating the Systems.

Beaumont \& Sofronoff \citep{Bea08} designed a CAS named \textit{``Junior Detective Training Program''} which included a detective themed storyline along with goals and rewards at the succession of a task which accumulate at the end. They increased levels of difficulty as individual move ahead in the CAS. Individuals were asked to identify emotional responses and then label correct responses for the characters in multiple settings. Twenty four adolescents with ASD were included in the study while the intervention group demonstrated an increase in pre-assessment to post-assessment results. 

Sansosti et al. \citep{San08} used spatio-temporal evolution of images as \citep{Sim04} \citep{Nik07} CAS in which social skills were taught to the individuals like joining in, greetings and sharing. The CAS involved video modeled “social stories” to illustrate correct social behaviors. Three individuals with HFA were included in the study. The authors reported CAS combined with spatio-temporal evolution of images to increased frequency of the targeted behaviors. Two week intervention was applied to the individuals within an educational environment and the follow up after showed that each individual was able to maintain the learned targeted behavior / communication skill.

Spatio-temporal evolution of image or video modeling isn't actually CAS / CVAT but an essential part of it, which is why it is included in this review to represent it as an initial step towards development of an efficient CAS / CVAT. Video modeling has been utilized to expand recurrence and diversity of social practices, and in a perfect world a significant part of a child's social repertoire should comprise of non-single social play, i.e. play including no less than one individual other than self. Most of these examinations show proof that video modeling can build or train a person of some target social skills. Target practices for treatment will be chosen as, for the person's shortages, including marking of feelings, free play, unconstrained welcome, talked dialects appreciation, conversational discourse, co-agent play, day-by-day living aptitudes and social play.

Grynszpan et al. \citep{Gry08} developed a CAS in a pre-post evaluation design. The first modality of the pre-design phase of the CAS called ``What to Choose?'' prompted the player to choose one of the three assertions about a dialogue scenario from which one assertion would be true e.g. Carole had an exam today and the teacher is away. A friend told Carole that the exam is canceled. How would Carole feel about this? A facial expression of a human like character is also put up with the dialogue so as to make it easier to determine how Carole is feeling regarding the scenario / dialogue. The characters were based on Ekman's rules \citep{Ekm03} designed with \textit{Poser}. The second modality of the CAS called ``Cartoon Characters''. Cartoon characters was similar to the previous one but the human like characters were replaced with cartoonic characters in this modality to prove which one would be considered or favored. Though  it did not seem to produce any differences in the performance. While the third module named ``Faces'' was also developed which was to solely recognize the facial expressions. A facial expression image was displayed along with 6 possible choices i.e. ( \textit{Happy, Sad, Surprised, Angry, Frightened, and without emotion}) and the player had to choose one of the six emotional expressions stated. On the other hand, post-design CAS presented a similar pattern, and called it ``intruder'', but with different content. The post-evaluation assessed participants capabilities of learning transfer and the acquired skills via training to a more complex task, including similar cognitive skills. As emphasized by Loveland \citep{Lov05}, the power to associate perceived emotions with a non-still social context is considered impaired in autism.

The training lasted for a period of three months, consisting of thirteen sessions. Two groups were made which were matched by age and academic level. A clinical group which included ten teenage boys with HFA and having an average chronological age of 12.10 years. A control group which comprised of normal typically developing individuals served as a reference base for the clinical group. The average IQ of HFA individuals with the WISC [87] (Wechsler Intelligence Scale for Children) was 80.5 . Results after evaluation of participants with autism theorized that their learning transfer was not so simple compared to a simple and a rich interface. Whereas the post assessment results demonstrated that the clinical group improved only with the simple interface. Thus, the rich interface must have hampered learning transfer for the clinical group. 

Lacava et al. \citep{Lac10} taught four children with HFA to identify the basic to complex emotions and to measure their social behavioral skills resulting in positive outcomes i.e. improvements in the recognition but without any quantitative conclusions. Whalen et al. \citep{Wha10} conducted a large scale study on 47 individuals with ASD aged 3-6 years. The study reported improvements in receptive language (object recognition) and social skills (expression recognition and perception) after eight weeks of training with ``TeachTown''- CAS. In order to measure potential of CAS for learned social skills, Hopkins et al. \citep{Hop11} tested their intervention named ``FaceSay'' CAS with a group of HFA and LFA. Forty-nine individuals (24 HFA \& 25 LFA) having IQ of 91.9 and 55.1 respectively were involved in ``FaceSay'' CAS intervention that targeted social skills. Individuals with ASD aged 6-15 completed twelve sessions over the period of six weeks with the CAS. Corresponding to the control group the intervention group showed improved social interactions and behaviors with peers which was observed while they interacted with each other in a playground.

\subsection{Virtual Reality / Augmented Reality}\label{VR}

Virtual Reality (VR) provides a framework to simulate real world scenarios using computer vision and computer graphics. Thus, VR technology allows instructors and therapists to offer a safe, repeatable and diversifiable environment during learning  for individuals facing ASD \citep{Par04}. Artificial Intelligence (AI) inculcates human-like senses and provides realism alongside stimulus or environmental control, which facilitates scenario creation required for learning in ASD intervention / therapy. VR advances have demonstrated potential for learning and evaluation of kids, young people, and grown-ups with a mental imbalance.


In the last decade, Moore et al. \citep{Moo05} used VR as a CVAT changing the technological medium of CAS for ASD. \citep{Moo05}  used emotional facial expression recognition with an anthropological 'avatar' for individuals of ASD. The CVAT allowed representation of facial expressions and animated sequences conveying emotional scenarios. The avatar displayed four exclusive expressions happy, angry, sad and afraid. The study was conducted in modular fashion. In the first module, individuals were asked to recognize expression of computer generated avatar. Second module was expression prediction. While, third module was designed to test / enhance individuals skill in predicting event that causes different expressions. Analysis of the data was done by relatively comparing the total observed responses to the total responses expected if they were selected by chance. The author reported that 90\% of the individuals who participated, were able to recognize, predict and interpret the avatar's emotional expressions. Moreover, they concluded that this CAS methodology allowed the children to effectively communicate with others.

Parsons et al. \citep{Par06} measured the behavior of two individuals (adolescent) with ASD, with the help of different VR settings, a cafe and a bus. In the study, Parsons et al. verified that adolescents considerably interpreted the scenes and appreciated the opportunities to keep up a dialogue. They also answered properly, although they indicated repetitive behavior and interpreted the things virtually. Mitchell et al. \citep{Mit07} followed a similar mode of analysis, by developing a virtual cafe. Half a dozen adolescents with ASD aged 14-16 having a mean Verbal IQ (VIQ) of 81.9, Performance IQ (PIQ) of 87.1, and full scale IQ (FSIQ) of 83.1 (measured by WASI \citep{Wec99}), were shown set of videos of real life scenarios going down in a bus or cafe in a virtual environment. The participants had to give explanation of wherever they had set to sit down and why e.g.by pointing towards a seat in a bus and as to why that seat? The virtual environment was developed with \textsl{Superscape Virtual Reality Toolkit} \citep{Zen98}. It was executed on a laptop using visualizer software presented in \citep{Par04}. To observe the VR environment, a joystick or a mouse was used. While to select object or person if one needed to communicate with a person in the VR environment was done using a click of the mouse. All the sessions were recorded for later analysis and evaluation. The training lasted for over six weeks. The researchers theorized that there have been cases that judged improvements, associated with time spent within the VR setting once deciding and explaining wherever they selected to sit on the bus. 

Fabri et al. \citep{Fab07} based their study on how individuals with ASD interact with characters (avatars) which are capable of showing facial expressions like (happiness, sadness, anger and fear). Within the first stage of the experiment, the participants, who were thirty-four youngsters with ASD, having mean-age of 9.96 had to choose the expression which was displayed by the avatar. In the second stage, individuals had to interpret what emotion the scene induced while the avatar mimicked it. In the third step, the individuals needed to choose what it was that caused the expression the avatar was expressing. The authors verified that thirty participants figured out the emotions of the character and utilized them suitably. While the remaining four participants, had severe autism.

While Ke \& Im \citep{Ke13}, worked with virtual reality environments involving  individuals with ASD in social scenarios. The primary task was to recognize the visual communication and facial expressions of characters present. The second was to interact with them in a college canteen. Lastly, interact with them at a party. \citep{Ke13} allotted associative analysis based on perception of the individuals and the questionnaires. They obtained positive results because the kids demonstrated that communication and interaction throughout the intervention had redoubled as did their communicative abilities after performing the tasks.

Strickland et al. \citep{Str13} devised a CVAT known as \textit{``JobTIPS''} that made it attainable to show employment especially interview related skills for adult individuals suffering from mild ASD or HFA. They used visual aids, guides, videos on the theory of mind \citep{Bar97} and virtual world \citep{Str13}. Twenty-two youngsters were engaged in the experiment to assess the efficiency of the CVAT. Half of the individuals completed sessions whereas the other half made up the control group. The control group didn't use the CVAT. Post- intervention, the participants who used the CVAT showed considerably higher verbal skills during the interview than the control individuals. 

Kandalaft et al. \citep{Kan13} designed a VR intervention to improve social cognitive skills of adults with ASD. There were eight participants with mean age of 21.25 and full scale IQ (FSIQ) of 111.88. The individuals practiced scripted role-playing scenarios during an interaction virtually with an online multi-player entertainment game along with two facilitators (clinicians) to help, if needed. The CVAT was designed by using Second Life v2.1 (Linden Lab 2003), a 3D virtual environment software available for public use. The CVAT included locations like office building, pool hall, a fast food chain, a tech store an apartment and so on. While the training scenarios or skills targeted were social interaction with a friend, initiating a conversation in a room, meeting strangers, negotiating in a store, giving a job interview, consoling a friend, blind dates, etc. majorly targeting verbal and non-verbal skills. The avatars representing the users in the VR world were modeled to resemble the participants and were driven by standard QWERTY keyboard. All participants were trained with 10 sessions of VR training and showed slight improvements but improvements were observed in all the targeted skill sets and most of the scenarios. 

Serret et al. \citep{Ser14} designed a slightly different CAS called \textit{``JeStimule''} which targeted facial expression and gesture recognition through visual non-verbal codes which were presented in the form of colors and symbols. Each basic emotion was associated with a color from Plutchik's emotional wheel \citep{Plu01} \textit{(happiness = yellow, anger = red, disgust = purple, fear = green, sadness = light blue, surprise = dark blue)}, while pain and neutral were associated with black and white respectively, based on a multi-senory virtual reality environment. The CAS was particularly designed for individuals facing LFA as there is no verbal text that needs to be read and only controlled via color codes that are associated with emotions for example red for anger. There were three levels of recognition that exist: recognizing the emotion expressed by a virtual character due to a specific event (e.g. 
a child falls down), the same task but the face of the virtual character is not revealed, recognizing the emotion conveyed by the non-verbal  communicative behavior of a character speaking with another character in which the verbal exchange is also inaudible. No conclusion and inferences from the statistical studies were concluded that showed positive results or improvements generically.

Cassidy et al. \citep{Cas15} worked to model ASD individuals (adults) perception of emotional expressions. Authors experimented with seventeen adults and adolescents (two females and fifteen males, aged 14-21 years having VIQ of 89.6, PIQ of 97.0 and FSIQ of 92.2). Individuals were assessed using  two conditions, i.e. dynamic and static. The CVAT included twenty-one video-clips of people showing facial expressions while receiving a present (a box of chocolates, monopoly money, etc). While the static stimuli were gained by a single frame taken from each dynamic stimuli were shown on the Tobii 1750 eye tracker. Tobii studio was used to record eye movements. Complete details of the study can be obtained from \citep{Cas14}. Results showed improvements in individuals with ASD in static conditions while not so significant in dynamics as the stimuli is fast paced and individuals were facing difficulty answering correctly.

\section{Summary} \label{Sum}

\begin{figure}[htb]
\includegraphics[scale=0.50]{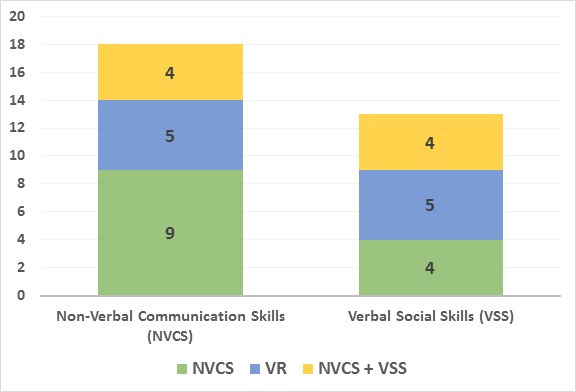}
\centering
\caption{Graphical representation of number of reviewed studies for distinct intervention techniques. NVCS depicts number of studies that target non-verbal communication skills, VSS depicts number of studies that target verbal social skills, while VR represents number of studies in NVCS and VSS that utilizes virtual reality technology.}
\label{fig1}
\end{figure}

This section presents most influential / innovative reviewed studies in tabular form (Table \ref{Tab}). The resulting 31 studies (graphically presented in Figure \ref{fig1}) include a total of 610 participants from which 550 individuals had ASD, while the remaining 60 participants were typically developing individuals. The reviewed 31 studies presented different intervention skills which are categorized into Non-Verbal Communication Skills (NVCS) and Verbal Social Skills (VSS) as described above, refer Section \ref{FEBI}.

The review for non-verbal communication skills included 18 studies  \citep{Tan10} \citep{Tsa11} \citep{Coc08}  \citep{Sil01} \citep{Gry08}  \citep{Boe02} \citep{Moo05} \citep{Gol06} \citep{Lac07} \citep{Lac10} \citep{Faj07} \citep{Gol09}   \citep{Hop11} \citep{Faj12}  \citep{Kan13} \citep{Ser14} \citep{Cas15} \citep{Gry07} from which 4 studies \citep{Gol06} \citep{Lac07} \citep{Lac10} \citep{Kan13}  were a combination of both non-verbal and verbal social skills (NVCS + VSS). 

While verbal social skills included 13 studies \citep{Ber01} \citep{Wha10} \citep{Str13} \citep{Bea08} \citep{Col73} \citep{Wha06} \citep{Gol06} \citep{Lac07} \citep{Lac10} \citep{Sim04} \citep{Par06} \citep{Mit07} \citep{Ke13} \citep{Kan13} i.e. some are combination of non-verbal communication which is defined in table \ref{Tab}. Moreover, a total of 09 studies from which 5 studies \citep{Str13} \citep{Par06} \citep{Mit07} \citep{Ke13} \citep{Kan13} are the inculcation of VR / AR into CAS / CVAT with respect to verbal social skills while 4 others \citep{Gry08} \citep{Moo05} \citep{Ser14} \citep{Cas14} are the inculcation of VR / AR into CAS/ CVAT with respect to non-verbal social skills.

Table \ref{Tab} defined below has the fields described as:
\begin{itemize}
\item Author(s) / Year published.
\item Mean Age of participants with (SD) standard deviation. 
\item Average IQ of individuals [SD]. 
\item Participant's characteristics.
\item Skills that were intervened by the CAS.
\item Type of Computer Aided System (CAS) used for the intervention.
\end{itemize}
The abbreviations that are used in the summarized Table \ref{Tab} are: 

\textit{TYP = Typically Developing individuals; N/A = Not Available; ASD = Autism Spectrum Disorder; SD = Standard Deviation; Avg = Average; HFA = High Functioning Autism; LFA = Low Functioning Autism; AS = Aspergers Syndrome; PDD-NOS = Pervasive Developmental Disorder Not Otherwise Specified; NS = Not Specified; NR = Not Reported; N = No.of individuals}


         \begin{landscape}

					
          \begin{longtable}{L{2.5cm}L{2cm}L{2cm}L{1.5cm}L{3.5cm}L{3cm}L{5cm}L{3cm}}
					\caption{Summary \& Analysis of Reviewed Studies}
          \\ \hline

\textbf{Author (s)}                                                                 
& \textbf{Participants}                                         
& \textbf{Mean Age (SD)}                                                           
 & \textbf{Avg. IQ {[}SD{]}}                                  
 & \textbf{Participant characteristics}                                                                                                                                 
 & \textbf{Skills intervened}
  & \textbf{Technology used for intervention}   
\\ 
\hline \\
\endhead
\multicolumn{3}{@{}l}{\ldots \textit{continued on next page}}

\endfoot

\endlastfoot
Colby \citep{Col73}            
  & 17 individuals with ASD                                      
	& N.R                                                                            
		& NR                                                                                                    
			& ASD (non speaking)                                                  
																																																																									& Generating voluntary speech for social interaction.  
																																																																							
																																																																							  & Keyboard-controlled audio-visual output.

																																																																						     \textbf{Results:} 13 individuals were reported increase in voluntary speech in social communication. Pioneering work.

\\ 
\hline
\\
Bernard-Opitz et al. \citep{Ber01}
2001                                                      
 & 16 (8 TYP, 8 ASD)                                             
 &  TYP 4 years ; ASD 7 years(1.1)         
 & 90                                                                                                    
  &  Eight individuals with mild to moderate ASD (2 females and 6 males). Eight individuals typical control (3 females and 5 males).                                   
	&    Social skills, Generating solution for conflicting. scenarios.                                                                                                
	&  \textbf{\textit{Software}:} Author developed application.
	
	    \textbf{\textit{Hardware}:} Personal Computer with windows 95. 
	
			\textbf{\textit{Setting}:} NS
			
			\textbf{\textit{Time}:} 10 sessions.
			
			\textbf{\textit{Results}:} Compared computer and instructor taught sessions on an enthusiasm scale. Reported lesser behavior issues and improved motivation in individuals using CAS. Treatment fidelity was not reported.
			& 
			
			
\\ 
\hline
\\

Silver \& Oakes \citep{Sil01}
2001
&22 ASD/AS	
&10-18 years	
&N/A	
&Eleven individuals were trained with CAS while eleven individuals were in control group(no intervention).
&Non-verbal communication skills, recognition of emotions from facial images, animated cartoon characters depicting situations.	
&\textbf{\textit{Software}:} Emotion Trainer.

\textbf{\textit{Hardware}:} NS

\textbf{\textit{Setting}:} School / Classroom.

\textbf{\textit{Time}:} 10 Sessions of 30 minutes for 2-3 weeks.

\textbf{\textit{Results}:} Improved recognition of facial expressions, in cartoonic characters displaying emotions and in story modes. Suggestive results. Treatment fidelity was not reported. Two out of three modalities depicted statistically significant results. 

																																					                    \\ 
\hline
\\
Bolte et al. \citep{Boe02} 
2002                                                                 
 & 10 AS/HFA                                                     
& AS/HFA 27.2 years (7.0)                                                             
& 104.2

{[}17.1{]}                                                                                        
 & 10 AS/HFA Male individuals from which N = 5 were intervened with CAS while others were assessed without intervention.                                                                                             
& Non-verbal communication skills, facial affect recognition.                                                                                               
 &\textbf{\textit{Software}:} Research developed application.

\textbf{\textit{Hardware}:} NS

\textbf{\textit{Setting}:} NS	

\textbf{\textit{Time}:} 2 hour training session per week for 5 weeks.

\textbf{\textit{Results}:} Quantitative conclusions were not possible due to missing information. Statistically consequential differences were found comparatively in control and treatment groups. Especially in assessment of reading the mind in the eyes and reading the mind in the face.      

\\ 
\hline
\\
Simpson et al. \citep{Sim04}
2004
	&4 ASD	&5.5 years (1.0)	
	&N/A	
	&4 participants with mild to moderate ASD (two males and two females).	
	&Social skills, spontaneous greetings to peers.	
	&\textbf{\textit{Software}:} HyperStudio 3.2
	
	\textbf{\textit{Hardware}:} PowerMac 5300.
	
	\textbf{\textit{Setting}:} A classroom
	
	\textbf{\textit{Time}:} 24 days (45 minutes per session)

	\textbf{\textit{Results}:} Individuals depicted improvement in the targeted social skill though no individual modality assessment was made to show which modality was more effective. 
	
	
\\ 
\hline
\\
Moore et al. \citep{Moo05}
2005
	&34 ASD	&7-16 years (9.96)	
	& NS	
	&Individuals participated were 29 boys and 5 girls.	
	&Non-verbal communication skills, identifying facial expression recognition \& making inferences based on expressions.
	&\textbf{\textit{Software}:} Researcher developed
	
\textbf{	\textit{Hardware}:} NS
	
	\textbf{\textit{Setting}:} Home
	
	\textbf{\textit{Time}:} NS
	
	\textbf{\textit{Results}:} Results indicated 90\% of the individuals were able to recognize the expressions of the avatar because of repeated practice in the VR environment. Improved social communication.
\\ 
\hline
\\

 Golan and Baron Cohen \citep{Gol06} - Experiment 1

2006  
&  65 (41 ASD, 24 TYP)  
&  HFA/AS 30.5 years(10.3); TYP 25.3 years(9.1)  
&  HFA/AS: V.IQ(108.3) P.IQ(112.0);

 TYP: V.IQ(115.8) P.IQ(112.5)  
& 41 AS/HFA individuals from which two groups made of N = 19 that went through the CAM intervention and N = 22 were assessed without intervention. 24 TYP individuals were also assessed with CAM intervention. 
& Non-verbal communication skills, social skills.
 &  \textbf{\textit{Software}:} Cambridge Mind reading.

\textbf{\textit{Hardware}:} IBM Laptop.

 \textbf{\textit{Setting}:} Home. 

 \textbf{\textit{Time}:} 10 to 15 weeks (2 hours/week).

\textbf{\textit{Results}:} No conclusive statistically convincing results for (a) reading the mind in the eyes; (b) reading the mind in the voices; (c) reading the mind in the films.

\\ 
\hline
\\
 Golan and Baron Cohen \citep{Gol06} - Experiment 2 

2006
 & 65 (41 ASD, 24 TYP)                                           
 &  HFA/AS 25.5 years(9.3); TYP 25years(9.0)     
&  HFA/AS: V.IQ(105.7)
   P.IQ(103.9); 
	
	TYP: V.IQ(109.2),
	P.IQ(106.3)
  & 26 AS/HFA individuals from which two groups made of N = 13 that went through the CAM intervention and N = 13 were assessed without intervention, while 13 TYP individuals were also assessed with CAM intervention. 
	& Non-verbal communication skills, social skills.
	&  \textbf{\textit{Software}:} Cambridge Mind reading.
	
	\textbf{\textit{Hardware}:} IBM Laptop.
	
	\textbf{\textit{Setting}:} home.
	
	\textbf{\textit{Time}:} 10 weeks (2 hours/week) + weekly tutor sessions in groups.

	\textbf{\textit{Results}:} No conclusive statistically convincing results for (a) Face sub-test; (b) Voice sub-test; (c) concepts recognized sub-test.

\\ 
\hline
\\
Parsons et al. \citep{Par06}
2006
& 2 ASD (John and Mike) 

&	John: 14 years and Mike: 17.7 years
& John VIQ 70; PIQ 83; FSIQ 73
	
& Mike VIQ 91; PIQ 107; FSIQ 100.
&	Social skills, virtual reality 
&\textbf{\textit{Software}:} Researcher developed application with Superscape v5.5 (Superscape visualizer)

\textbf{\textit{Hardware}:}Laptop 650 MHz/64 MB RAM / Windows 98.
\textbf{\textit{Setting}:} NS.

\textbf{\textit{Time}:} 4 weeks.

\textbf{\textit{Results}:} Suggestive results. Qualitative. Positive outcome for engagement with CAS / CVAT. No empirical results were concluded.
 \\
\hline
\\ 
Faja et al. \citep{Faj07}

2007
& 10 HFA
& 19.6 years (7.4)
&99.4 

[13.9]
	&10 males with HFA 12-32 years of age. 
	&Non-verbal communication skills, facial recognition, immediate and delayed facial memory, categorization of faces based on age, group and individual identity.
	&\textbf{\textit{Software}:} E-Prime 1.0 experimental software, MS PowerPoint \& Adobe Photoshop for editing images.

\textbf{\textit{Hardware}:} laptop resolution at 1024x768 image resolution at 72 PPI.

\textbf{\textit{Setting}:} NS.

\textbf{\textit{Time}:} 8 training sessions lasting 30 minutes to 1 hour during a 3-week period.

\textbf{\textit{Results}:} Sensitivity to second order relation only depicted statistically significant results but not on six other measures of face processing. Treatment fidelity was not reported.
\\ 
\hline
\\ 

Lacava et al. \citep{Lac07}
2007
&8 AS	
&10.3 years (1.2)	
&N/A	
&8-11 years of age (6 males and 2 females) with AS.
&Non-verbal communication skills, social skills.	
&\textit{\textbf{Software:}} Cambridge mind reading.

\textbf{\textit{Hardware}}: Personal computer.

\textbf{\textit{Setting}:} Home and school.

\textbf{\textit{Time}:} 10 weeks (variable of individual sessions).

\textbf{\textit{Results}:} significant pre to post intervention results were reported in CAM-CAS
\\ 
\hline
\\

Mitchell et al. \citep{Mit07}
2007
& 7 ASD/AS 

&	14-16 years
&PIQ 87.1; VIQ 81.9; FSIQ 83.1	
& 7 Individuals with ASD/AS (four males and three females)
&	Social skills, virtual reality.
&\textbf{\textit{Software}:} Researcher developed application with Superscape Virtual Reality Toolkit

\textbf{\textit{Hardware}:} NS.

\textbf{\textit{Setting}:} NS.

\textbf{\textit{Time}:} 6 weeks.

\textbf{\textit{Results}:} Improved judgment and reasoning decisions. pre post VR intervention demonstrated little signs of gains.
\\ \hline
\\
Beaumont et al. \citep{Bea08}
2008
&49 AS
&7-11 years 	
&85 or above.	
&44 males and 5 females. Randomly assigned N = 26 to intervention while N = 23 to wait list groups.
&Social skills, non-verbal communication skills. 	
&\textit{\textbf{Software:}} Junior Detective training.

\textbf{\textit{Hardware}:} NS	

\textbf{\textit{Setting}:} Laboratory

\textbf{\textit{Time}:} 7 weeks.

\textbf{\textit{Results}:} statistical improvements were depicted. though no detailed description of hardware. Other intervention modalities may have contributed.


                                                                                                                                                                                           \\ 
\hline
\\
Cockburn et al. \citep{Coc08}
2008
&\textit{No experiment conducted.}

&-	
&NS	
&Individuals with absolute or  a relative impairment in face processing 
&	Non-verbal Communication skills, facial expression recognition.
&\textbf{\textit{Software}:} Smile Maze, researcher developed application

\textbf{\textit{Hardware}:}NS.

\textbf{\textit{Setting}:} NS.

\textbf{\textit{Time}:} NS.

\textbf{\textit{Results}:} no experiments were done.

\\ 
\hline
\\
Grynszpan et al. \citep{Gry07} \citep{Gry08}

2008  
   & 20 (10 TYP, 10 HFA)

	& TYP 9 years (1.0); HFA 12 years (1.2).                                                  
	& 80.5
	
	{[}18.5{]}.                                                                                        
	& Ten individuals with HFA (10 teenage boys). Ten individuals typical control (8 boys and 2 girls).
	& Non-verbal communication skills, social skills, virtual reality.
	&  \textbf{\textit{Software}:} Two, Author developed applications (training game and evaluation game)
	
	\textbf{\textit{Hardware}:}NS.
	
	\textbf{\textit{Setting}:} School.
	
	\textbf{\textit{Time}:} 3 months (11 training sessions).

\textbf{\textit{Results}:} ASD Individuals improved in the basic mode of the CAS while the control group showed improvement in both the modalities.
	

\\ 
\hline
\\
Golan et al. \citep{Gol09}
2009	
& 56 (38 HFA/AS; 18 TYP) 	
&TYP 5.4(1.1); ASD intervention 5.6(1.0); ASD control 6.2(1.0)	
&N/A	
&56 participants (N = 20 ASD participants were intervened with CAS; N = 18 ASD were control and baseline correlation with N = 18 TYP.)
&Non-verbal communication skills, emotional vocabulary.
&\textbf{\textit{Software}:} The Transporters DVD.

\textbf{\textit{Hardware}:} IBM PC with a 17 inch monitor.

\textbf{\textit{Setting}:} Testing Room/ Home.

\textbf{\textit{Time}:} 2-3 episodes per day for 4 weeks.

\textbf{\textit{Results}:} statistical measures were used to find treatment fidelity though no concrete empirical conclusions were made. However when correlations were calculated separately for each group positive correlations were found.

\\ 
\hline
\\
Lacava et al. \citep{Lac10}
2010	
&4 HFA 	
&8.6 years (0.8)	
&N/A	
&7-11 years of age and text and computer literate.	
&Non-verbal communication skills, social skills.
&\textbf{\textit{Software}:} Cambridge mind reading 

\textbf{\textit{Hardware}:}  Personal computer.

\textbf{\textit{Setting}:} Classroom.

\textbf{\textit{Time}:} 1-2 hours/week for 7-10 weeks.

\textbf{\textit{Results}:} sample size too small for any conclusive results. The focus was to prove the superiority of CAS over human instructed sessions. Positive improvements however were reported from participants.


\\ 
\hline
\\
Tanaka et al. \citep{Tan10}
2010                                                                
& 79 ASD                                                        
& ASD 10.5 years (3.8)                                                                 
& 93.6{[}22.1{]}                                                                                         
 & 79 ASD individuals from which N = 42 were intervened with CAS while N = 37 were waiting list. participants                                                                                                          
 & Non-verbal communication skills, facial features and expression recognition with LFI developed application.                                                                     
 & \textbf{ \textit{Software}:} Let's Face It! (LFI).

 \textbf{\textit{Hardware}:} NS.

 \textbf{\textit{Setting}:} Home.

\textbf{\textit{Time}:} 20 hours participants were given free hand. The only constraint was to complete 100 minutes per week training.

\textbf{\textit{Results}:} No significant effects depicted of the LFI intervention. Only part / whole identifying modality depicted significant results. Hardware not specified in sufficient detail.


\\ 
\hline
\\ 

Whalen et al. \citep{Wha10}
2010	
&47 ASD	
&3.8 years (0.8)	
&N/A	
&22 participants were trained while 25 non trained participants with ASD.	&Social Skills.
&\textbf{\textit{Software}:} TeachTown:Basics.

\textbf{\textit{Hardware}:} NS.

\textbf{\textit{Setting}: }Classroom.

\textbf{\textit{Time}:} 12 weeks(20 minutes/session).

\textbf{\textit{Results}:} Significant improvement only in receptive vocabulary size measured using Peabody picture vocab test \citep{dunn1997m}.

\\
\hline
\\
Hopkins et al. \citep{Hop11} 
2011                                                              
& 49(24 HFA; 25 LFA)                                           
 &  HFA 10.05 years (2.3); LFA 10.3 years (3.3)     
&  HFA 91.9 {[}19.5{]}; LFA 55.1 {[}20.9{]}                        
 &  24  HFA individuals from which N = 13 individuals were assessed while, 25 LFA individuals from which N = 11 were assessed                                           
 & Non-verbal communication skills, facial features.           
 & \textbf{ \textit{Software}:} FaceSay CAS application

 \textbf{\textit{Hardware}:} NS

 \textbf{\textit{Setting}:} School with trainers

 \textbf{\textit{Time}:}10-25 minutes/session for 6 weeks (12 sessions)

\textbf{\textit{Results}:}Individuals with HFA depicted improved in all areas of assessment while the individuals with LFA depicted improvements in emotion recognition and social interactions comparatively. LFA individuals had lower congnitive functioning.

\\ 
\hline
\\

Tsai et al. \citep{Tsa11}

2011
&\textit{No experiment conducted.}

&-	
&NS	
&Individuals with an absolute or relative impairment in face processing 
&	Facial expression recognition, non-verbal communication skills.
&\textbf{\textit{Software}:} Face Flower, researcher developed application.

\textbf{\textit{Hardware}:}NS.
\textbf{\textit{Setting}:} NS.

\textbf{\textit{Time}:} NS.

\textbf{\textit{Results}:} no experiment conducted.


\\ 
\hline
\\

Faja et al. \citep{Faj12}

2012
& 18 ASD	
& Face training
 22.4 years (4.4);
House Training 21.5 years (5.6)	
& Face training
116.3 [16.3];
House training 118.2 [17.4]	
& individuals were randomly assigned to either face training ( n = 9, 6 with absolute impairment) or to house training (n = 9; 7 with absolute impairment). 
&	Non-verbal communication skills.	
&\textbf{\textit{Software}:} E-Prime 1.0 experimental software, MS PowerPoint \& Adobe Photoshop for editing images.

\textbf{\textit{Hardware}:} laptop resolution at 1024x768 image resolution at 72 PPI.

\textbf{\textit{Setting}:} NS

\textbf{\textit{Time}:} 8 sessions or when criteria was met.

\textbf{\textit{Results}:} Authors reported improved memory of individuals assessed on either of the modalities : house / face recognition.


\\ 
\hline
\\
Jain et al. \citep{Jai12}

2012
&9 ASD (6 HFA;
2 LFA;
1 severe ASD)	
& 5 to 12 years	
& NS	
& Individuals with an absolute or relative impairment in face processing. 
&	Non-verbal communication skills, facial expression recognition, social skills by living in a game story.
&\textbf{\textit{Software}:} Researcher developed application.

\textbf{\textit{Hardware}:}NS.

\textbf{\textit{Setting}:} Homes or coffee shop.

\textbf{\textit{Time}:} As much as the participants want to in a single sitting.

\textbf{\textit{Results}:} no empirical statistical results were concluded. Claims improved recognition in individuals.

                                                                                                \\ 
\hline
\\

Kandalaft et al. \citep{Kan13}
2013
& 8 ASD / AS 

&	21.25 years (2.71)
&FSIQ 111.88 [8.51]	
& 8 Individuals with ASD / AS (six males and two females)
&	Social skills, non verbal skills, virtual reality 
&\textbf{\textit{Software}:} Researcher developed application with \textit{Second Life} v2.1 (Linden Lab 2003)

\textbf{\textit{Hardware}:} ATi Radeon 8500/ 1.5 GHz CPU with 24 inch 1920x1200 monitor.

\textbf{\textit{Setting}:} NS.

\textbf{\textit{Time}:} 10 sessions.

\textbf{\textit{Results}:} Individuals depicted some improvements in emotion recognition and the ability to infer the emotions in others in the form of increased conversing skills.

 \\
\hline
\\ 

Strickland et al. \citep{Str13}
2013
& 22 AS/HFA 

&	18.21 years (1.03)
&NS	
& 11 individuals received intervention while the other 11 did not go through intervention.
&	Teaching social skills, job interview skills, virtual reality 
&\textbf{\textit{Software}:} JobTIPS, researcher developed application.

\textbf{\textit{Hardware}:} NS.

\textbf{\textit{Setting}:} University room.

\textbf{\textit{Time}:} 30 minutes per session.

\textbf{\textit{Results}:} Pre to post test intervention depicted transfer of conversing skills in individuals.

 \\
\hline
\\ 
KE \& IM \citep{Ke13}

2013
& 04 AS / HFA 

&	9-10 years
& NS	

&  --
&	Teaching social skills, initiation, virtual reality 
&\textbf{\textit{Software}:} Researcher developed application.

\textbf{\textit{Hardware}:} NS.

\textbf{\textit{Setting}:} School, library.

\textbf{\textit{Time}:} 1 week (3 hours session)

\textbf{\textit{Results}:} Positive outcome of the individuals demonstrated that communication and interaction along the duration of the intervention increased as did their communicative competences after completing the tasks.

 
 \\
\hline
\\ 
Serret et al. \citep{Ser14}

2014
& 33 (23 HFA / LFA; 4 AS; 6 PDD-NOS)

&	11.4 years (3.16)
& NS	

& 33 individuals (31 males and 2 females from which 19 can read while 14 are non-readers).
&	Training emotion recognition skills, facial expressions, virtual reality
&\textbf{\textit{Software}:} JeStiMulE, researcher developed application. 

\textbf{\textit{Hardware}:} NS.

\textbf{\textit{Setting}:} A room.

\textbf{\textit{Time}:} 1-hour session twice a week for 4 weeks.

\textbf{\textit{Results}:} Results depicted suitable adaptability. Improvements were seen after sessions on avatars and on real life pictures of characters.

 \\
\hline
\\
Cassidy et al. \citep{Cas15}
2015
& 17 ASD 

&	14-21 years
&VIQ of 89.6 ; PIQ of 97.0 and FSIQ of 92.2	

& 17 ASD (15 males and 2 females)
&	Facial expressions, non-verbal communication skills, virtual reality 
&\textbf{\textit{Software}:} Researcher developed application.

\textbf{\textit{Hardware}: } Tobii 1750 eye tracker.

\textbf{\textit{Setting}:} NS.

\textbf{\textit{Time}:} As much as needed.

\textbf{\textit{Results}:} Individuals depicted improvements from the static stimuli compared to the dynamic stimuli.


\\\hline

%
\label{Tab}
\end{longtable}


TYP = Typically Developing individuals; N/A = Not Available; ASD = Autism Spectrum Disorder; SD = Standard Deviation; Avg = Average; HFA = High Functioning Autism; LFA = Low Functioning Autism; AS = Aspergers Syndrome; PDD-NOS = Pervasive Developmental Disorder Not Otherwise Specified; NS = Not Specified; N = No.of individuals; VR = Virtual Reality 
         

\end{landscape}

\section{Conclusion and Future Directions}

This survey describes how individuals of Autism Spectrum Disorders (ASD) are handled by technology based intervention methodologies, which include: contemporary Computer Aided Systems (CAS), Computer Vision Assisted Technologies (CVAT) and  Virtual Reality (VR). Individuals with ASD have difficulty interacting with their peers, teachers and parents, etc. Most of them prefer to interact with computers or augmented reality systems, which proves that technology based interventions can be an effective tool for enhancing the purpose of teaching and learning facial expressions as well as social skills. As summarized in Table \ref{Tab}, most of the technology based interventions are not designed to fulfill needs of individuals suffering from ASD completely, as they have variable needs.

Major conclusions and future directions drawn from this extensive literature survey are presented below:

\begin {enumerate}

\item Generally, surveyed literature concluded much promise for CAS / CVAT interventions although none of the studies reported concrete evidence that it was able to train or modify a complete behavior of the selected individuals. However, many studies reported significant improvements often through statistical validations. 

\item There is a need to design an intervention technology that follows a standard which is accepted to the clinicians as well as to the AI research community. An effort to standardize intervention technologies can be seen in a recent study in robotics \citep{Beg16}, while Clark et al. \citep{Cla05} \& Higgins et al. \citep{Hig96} suggested principles to guide the development of an effective CAS / CVAT.

\item There is a dire need to create standard audio / video database of individuals facing ASD. In the absence of such database making artificial intelligence / machine learning algorithms is highly challenging. Secondly, the efficacy or robustness of published results remain doubtful or uncertain.

\item Most of the researchers described their CAS / CVAT in a therapeutic or a clinical fashioned way. They demonstrated their findings and observations but do not elaborate the architecture and design of CAS / CVAT thoroughly i.e. the way an application is structured and what algorithms are defined and utilized in developing those CAS / CVAT w.r.t. computer vision (CV). While some researchers described and elaborated the architecture w.r.t CV domain \citep{Jai11} \citep{Tsa11} \citep{Coc08} \citep{Jai12}.

\item It is required to develop flexible intervention technology targeting the deficit skills in an individualized instruction settings. Designing an intervention technology, it is to be considered that the system design should handle the common problems and cater the generic issues faced by individuals of ASD. Thus, developing a single handed solution that can work with a variety of deficit issues which are described above in section \ref{FEBI}.


\item  Collaborative efforts are required by the research community of AI, psychology and neuroscience to develop an AI assisted intervention technology for individuals suffering from ASD, to better understand needs of both the worlds. Scientists should focus on to remove the barrier and boundaries between disciplines and move towards a broader perspective of education and contribution towards the scientific community by providing more essential and effective findings.

\end {enumerate}






\end{document}